\documentclass[useAMS,usenatbib]{mn2e}

\usepackage{amsmath}
\usepackage{amssymb}
\usepackage{soul}
\usepackage{natbib}
\usepackage{graphicx}
\usepackage{caption}
\usepackage{subcaption}
\usepackage{enumerate}
\usepackage{datetime}
\usepackage{color}
\usepackage{hyperref}
\usepackage{draftcopy}

\begin{document}

\title[Analysis of Multiple Resonances]
{Line-driven ablation of circumstellar discs: II. Analyzing the role of multiple resonances}

\author[N. D. Kee et al.]
{Nathaniel Dylan Kee$^1$\thanks{Email: nathaniel-dylan.kee@uni-tuebingen.de},
Stanley Owocki$^2$, and Rolf Kuiper$^1$\\
 $^1$ Institut f\"ur Astronomie und Astrophysik, Eberhard Karls Universit\"at T\"ubingen, D-72076 T\"ubingen, Germany\\
 $^2$ Department of Physics and Astronomy, Bartol Research Institute,
 University of Delaware, Newark, DE 19716, USA\\ 
 }

\def\<<{{\ll}}
\def\>>{{\gg}}
\def\wig{{\sim}}
\def\spose#1{\hbox to 0pt{#1\hss}}
\def\+/-{{\pm}}
\def\=={{\equiv}}
\def\mubar{{\bar \mu}}
\def\mustar{\mu_{\ast}}
\def\Lambar{{\bar \Lambda}}
\def\Rstar{R_{\ast}}
\def\Mstar{M_{\ast}}
\def\Lstar{L_{\ast}}
\def\Tstar{T_{\ast}}
\def\gstar{g_{\ast}}
\def\vth{v_{th}}
\def\grad{g_{rad}}
\def\glines{g_\mathrm{lines}}
\def\Mdot{\dot M}
\def\mdot{\dot m}
\def\yr{{\rm yr}}
\def\ksec{{\rm ksec}}
\def\kms{{\rm km/s}}
\def\qad{\dot q_{ad}}
\def\qlines{\dot q_\mathrm{lines}}
\def\solar{\odot}
\def\Msun{M_{\solar}}
\def\msbyr{\Msun/\yr}
\def\Rsun{R_{\solar}}
\def\Lsun{L_{\solar}}
\def\Be{{\rm Be}}
\def\Rpole{R_{\ast,p}}
\def\Req{R_{\ast,eq}}
\def\Rmin{R_{\rm min}}
\def\Rmax{R_{\rm max}}
\def\Rstag{R_{\rm stag}}
\def\vinf{V_\infty}
\def\Vrot{V_{rot}}
\def\Vcrit{V_{crit}}
\def\d{\mathrm{d}}
\def\rres{\mathbf{r}_\mathrm{res}}
\def\ro{\mathbf{r}_\mathrm{loc}}

\maketitle

\begin{abstract}
We extend our previous study of radiative ablation of circumstellar disks by line-scattering of the star's radiation, accounting now for the effect of multiple line resonances off the stellar limb. 
For an analytic, three-dimensional model of the velocity structure of an equatorial Keplerian disk bounded at higher latitudes by a radially accelerating stellar wind outflow, we use root-finding methods to identify multiple resonances from a near-disk circumstellar location along starward rays both on and off the stellar core.
Compared to our previous study that accounted only for the effect of on-core resonances in reducing the radiative driving through the scattering of radiation away from a near-disk circumstellar location, including off-limb resonances leads to additional radiative driving from scattering toward this location.
Instead of the up-to-50\% reduction in line-acceleration previously inferred from multiple resonance effects, we now find a more modest 15-20\% net reduction.
\end{abstract}

 
\section{Introduction}
\label{sec:intro}

For a radially expanding wind, the line-of-sight velocity is monotonic along all lines of sight.
This means that a photon Doppler shifted into resonance with a given spectral line scatters, and then escapes the system without being Doppler shifted into resonance with the same spectral line again.
This fact is at the heart of the purely local calculations of line-acceleration in the \cite{CasAbb75} (hereafter, CAK) theory.
However, one the velocity along a line-of-sight becomes non-monotonic, photons traveling from the stellar core can be Doppler shifted into resonance, scatter onto one of these lines of sight where the line-of-sight velocity is non-monotonic, and then be Doppler shifted into resonance with the same line a second time.
The net result of these ``multiple resonances" is that calculations of line-acceleration cannot use the unattenuated stellar intensity, and thus become inherently non-local.

For one-dimensional and simple two-dimensional geometries, it is possible to approximate, or even directly account for,  this non-locality in hydrodynamic simulations.
For example, take the one-dimensional radiation hydrodynamics simulations of \cite{FelNik06}.
By using a spherically symmetric wind with a non-monotonic radial velocity as an initial condition, they induced multiple resonances.
Using the source function in the presence of multiple resonances as derived by \cite{RybHum78}, \cite{FelNik06} found that a spherically symmetric wind that goes through a velocity reversal anywhere will continue to show a decelerating flow at all larger radii, eventually reaching a coasting solution with drastically reduced terminal speed.

Simulations of the line-deshadowing instability, or LDI \citep[see, e.g.][]{DesOwo03,DesOwo05,SunOwo13,SunOwo15} also treat multiple resonances, now by approximating the scattering source function, for instance with the smooth-source function method \citep[SSF:][]{Owo91} or the escape-integral source function method \citep[EISF:][]{OwoPul96}, and then integrating over long characteristics to solve for the line-acceleration at all locations.
Such simulations, due to their use of long-characteristics, are quite computationally expensive, and therefore have had to be limited to solving for the line-acceleration along one \citep[see, e.g.][]{DesOwo03} or three (\cite{DesOwo05}, Sundqvist et al., in prep) characteristics shared by multiple grid cells.

In complicated three-dimensional geometries, the line acceleration must be calculated along tens of rays unique to each cell.
Therefore, the majority of prior hydrodynamic simulations including line-acceleration in non-spherically symmetric environments \citep[see, e.g.][]{CraOwo96,ProSto99,udDSun13} have chosen to ignore multiple resonances.
This was also case for the majority of \cite{KeeOwo16} (hereafter paper I), where we introduced a mechanism for the ablation of disc material around massive (O and B type) stars by line-acceleration.
Already there, however, we identified that multiple resonances do arise in the inherently spherically asymmetric geometry of a circumstellar disc embedded in a stellar wind,
leading to the analysis presented in appendix A of paper I.
We used an analytic velocity structure representing such a wind+disc geometry, along with standard root-finding methods, to calculate that multiple resonances can lead to as much as a 50\% reduction in radial flux in the ablation layer. 

We here undertake an extension to the study in paper I, now accounting for the effects of off-star resonances and directly calculating the line-acceleration as opposed to calculating radial flux.
In section \ref{sec:mult_res}, we introduce the analytic wind and disc profiles used for this analysis, as well as providing an overview of the multiple resonances that arise in such a geometry.
In section \ref{sec:root-find}, we then introduce the methods used to analyze the effects of multiple resonances on line-acceleration.
Section \ref{sec:results} presents results of this analysis, and section \ref{sec:conclusions} summarizes the most important points of the paper, as well as presenting potential future research directions, both for analyses of multiple resonances and for studies of line-driven ablation.

\section{Model Geometry and Basics of Multiple Resonances}
\label{sec:mult_res}

For radially expanding winds, photons can only be Doppler shifted into resonance with a spectral line at one location,
at which point they scatter, and subsequently escape the system. For geometries where the line-of-sight velocity is not monotonic in all directions, however, photons can be scattered into directions where they will be Doppler shifted into resonance with the same spectral line again.
The presence of such multiple resonances breaks the locality of the calculation of line-acceleration as presented by CAK, resulting in the need for the line-acceleration to be calculated non-locally.

To illustrate this point with a concrete example, let us take the analytic velocity structure from the appendix of paper I, expressed as a function of the spherical radius $r$ and latitude\footnote{Note that this is in contrast to the \emph{co-}latitude used in standard spherical coordinates.} $\theta$. For the radial flow speed, $v_r$, we use the simple expression,

\begin{equation}\label{eqn:vr}
v_r(r,\theta)=V_\infty(\theta)\left(1-\gamma\frac{R_\ast}{r}\right)\;,
\end{equation}
in terms of the stellar radius, $R_\ast$, and the latitude dependent terminal speed, $V_\infty(\theta) \equiv v_r(\infty,\theta)$.
Note the introduction of $\gamma=1-c_s/V_{\infty,\mathrm{o}}$, a function of the terminal speed of the unperturbed wind, $V_{\infty,\mathrm{o}}$, and the isothermal sound speed, $c_s$.
Including this term keeps the radial velocity above or equal to $c_s$ in the wind and smoothly varying everywhere.

\begin{figure}
\centering
\begin{subfigure}{0.5\textwidth}
\includegraphics[width=\textwidth]{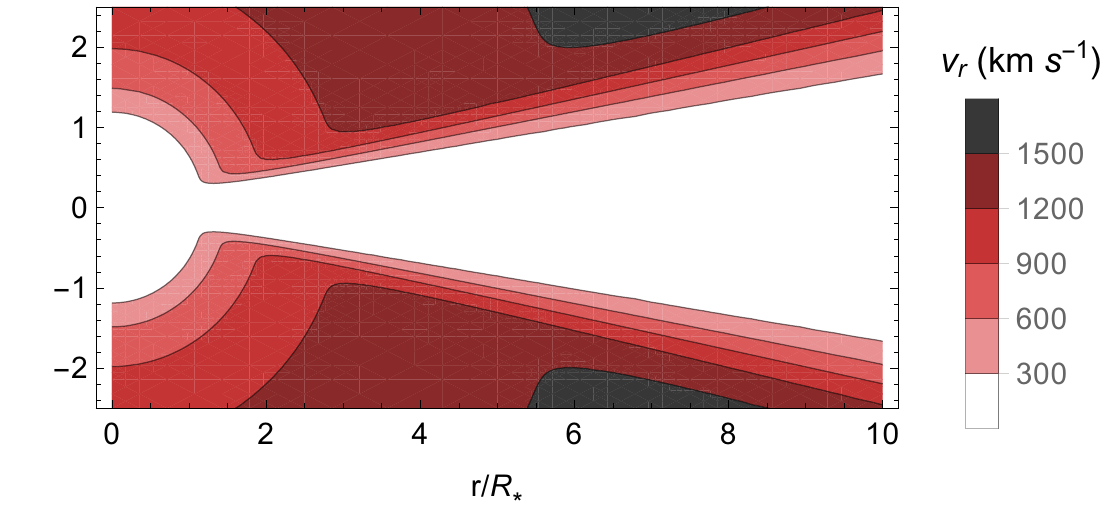}
\caption{$v_r$}
\end{subfigure}

\begin{subfigure}{0.5\textwidth}
\includegraphics[width=\textwidth]{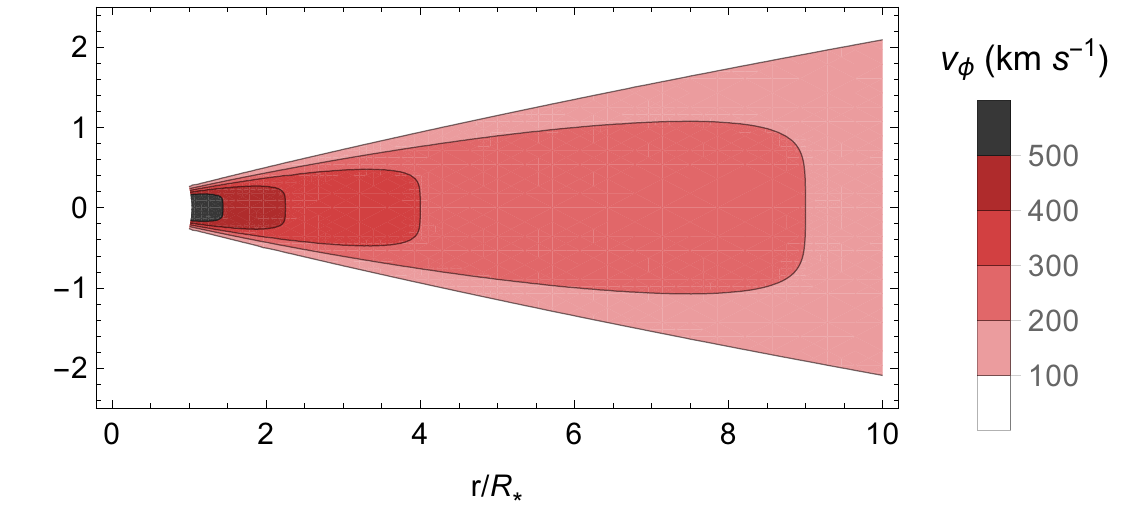}
\caption{$v_\phi$}
\end{subfigure}

\begin{subfigure}{0.5\textwidth}
\includegraphics[width=\textwidth]{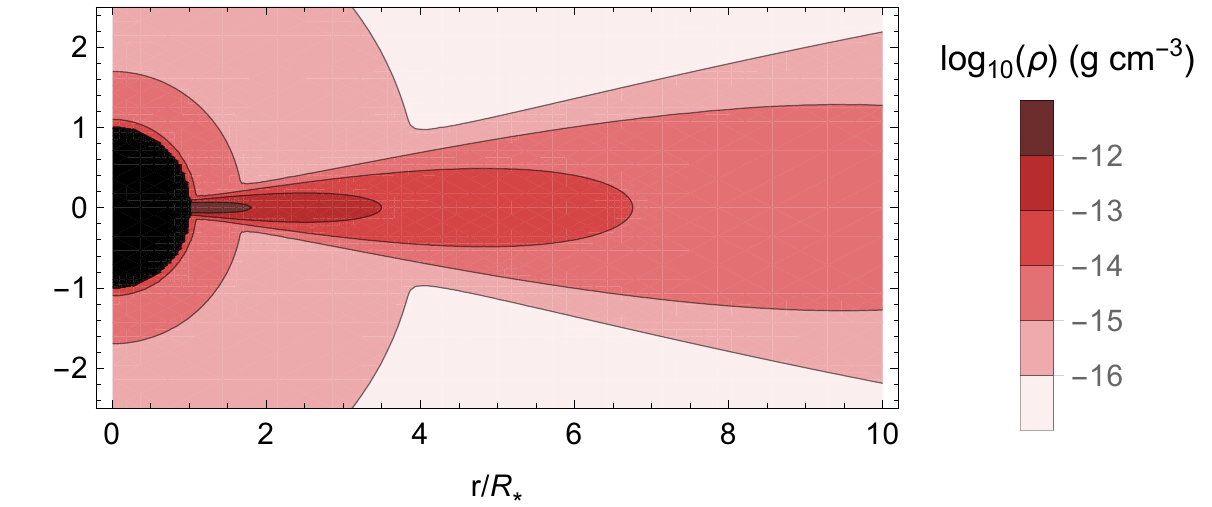}
\caption{$\log_{10}(\rho)$}
\end{subfigure}

\caption{From top to bottom a) radial and b) azimuthal components of velocity and c) density for $\tau_d=1$. Axes are plotted in units of stellar radii with contours for velocity in km s$^{-1}$ and density in g cm$^{-3}$.\label{fig:analytic}}
\end{figure}

To mimic the velocity structure associated with ablation, we choose the latitudinal variation $V_\infty(\theta)$ such that it returns the unperturbed wind terminal speed $V_{\infty,\mathrm{o}} \equiv V_\infty(90^\circ)$ over the poles, and decreases steeply around the disc-wind boundary to approach zero near the disc midplane.
Specifically, for a disc-wind boundary thickness of half width $\Delta \theta$ centered on latitudes $\pm \theta_d$, we use the complementary error function to give

\begin{equation}\label{eqn:vinf}
V_\infty(\theta)=\frac{V_{\infty,\mathrm{o}}}{2}\mathrm{Erfc}\left(\frac{\theta_d-\lvert \theta \rvert}{\Delta\theta}\right)\;.
\end{equation}
For the remainder of the paper, we use $\Delta \theta = 4^\circ$ and $\theta_{d}=12^\circ$ to match the asymptotic state of disk ablation found in paper I.
Panel (a) of figure \ref{fig:analytic} plots $v_r$ for $V_{\infty,\mathrm{o}}=1800$ km s$^{-1}$ and $c_s=20$ km s$^{-1}$.

For the azimuthal component of the flow speed, $v_\phi$, we use
\begin{equation}\label{eqn:vphi}
v_\phi(r,\theta)=V_\mathrm{orb}(\theta)\sqrt{\frac{R_\ast}{r\, \mathrm{Max}[\cos(\theta),10^{-3}]}}\;,
\end{equation}
where we introduce the angle dependence of the orbital speed as an extension to the appendix of paper I.
This angular dependence is given in terms of the orbital velocity on the equator at $r=R_\ast$, $V_\mathrm{orb,o}=V_\mathrm{orb}(0)=\sqrt{G M_\ast/R_\ast}$, itself a function of Newton's gravitational constant $G$ and the stellar mass $M_\ast$, to be

\begin{equation}\label{eqn:vorb}
V_\mathrm{orb}(\theta) \equiv v_\phi(R_\ast,\theta) = V_\mathrm{orb,o}\left[1-\frac{1}{2}\mathrm{Erfc}\left(\frac{\theta_d-\lvert \theta \rvert}{\Delta\theta}\right)\right]\;.
\end{equation}
Panel (b) of figure \ref{fig:analytic} plots $v_\phi$ for $V_{\mathrm{orb,o}}=600$ km s$^{-1}$.

Finally, as a further extension to the analysis from the appendix of paper I, we introduce an analytic density variation.
For the wind-like density profile desired at high latitudes, we select

\begin{align}\label{eqn:rho_w}
\rho_{w}(r,\theta)&=\frac{\dot{M}}{4\pi v_r(r,\theta) r^2}\\
&=\frac{\rho_{w,\mathrm{o}}V_{\infty,\mathrm{o}}}{(r/R_\ast)^2 v_{r}(r,\theta)}\;,
\end{align}
such that the wind mass loss rate $\dot{M}$ is independent of radius.
The second equality introduces the characteristic scale of wind density, $\rho_{w,\mathrm{o}}\equiv \dot{M}/(4\pi V_{\infty,\mathrm{o}}R_\ast^2)$.
For low latitudes, we select a disc-like density variation that assumes the disc is in hydrostatic equilibrium vertically with normalized scale height $h\equiv H/R_\ast=c_s/V_\mathrm{orb,o}$, and that the disc midplane density falls off as a power law of index $\beta$ in the cylindrical radial direction. 
Together this gives

\begin{equation}\label{eqn:rho_d}
\rho_d(r,\theta) = \rho_{d,\mathrm{o}}\left(\frac{r \cos(\theta)}{R_\ast}\right)^{-\beta}e^{h^{-2}\left(R_\ast/r-R_\ast/\left(r\cos(\theta)\right)\right)}\;,
\end{equation}
where the characteristic disc density\footnote{Note that this definition omits a factor of $(\beta - 1)$. This choice was made for consistency with Paper I, as well as to limit the dependence of the density scale to only the stellar radius and a characteristic optical depth.}, 
$\rho_{d,\mathrm{o}}\equiv\tau_d/\left(\kappa_e R_\ast\right)$, is defined additionally in terms of the midplane optical depth of the disc in the radial direction, $\tau_d$, and the continuum electron scattering opacity, $\kappa_e$.
In order to smoothly map from disc-like densities at low latitudes to wind-like densities at high latitudes, and for consistency with equations \ref{eqn:vr} and \ref{eqn:vphi}, we again use complementary error functions such that 

\begin{multline}\label{eqn:rho}
\rho(r,\theta)=\frac{\rho_w(r,\theta)}{2}\mathrm{Erfc}\left(\frac{\theta_d-\lvert\theta\rvert}{\Delta \theta}\right)\\
+\rho_d(r,\theta)\left[1-\frac{1}{2}\mathrm{Erfc}\left(\frac{\theta_d-\lvert\theta\rvert}{\Delta \theta}\right)\right]\;.
\end{multline}
Panel (c) of figure \ref{fig:analytic} plots $\rho$ for $\rho_{d,\mathrm{o}}=8\times10^{-12}$ g cm$^{-3}$ and $\rho_{w,\mathrm{o}}=1.2\times10^{-15}$ g cm$^{-3}$.

Taken together, these three-dimensional, analytic velocity and density structures allow us to also find an analytic variation of velocity and density along any arbitrary line-of-sight.
For simplicity, we normalize these line-of-sight velocities by $V_\mathrm{orb,o}$ and the densities along the line-of-sight by $\rho_{d,\mathrm{o}}$ in the remainder of the paper.
In order to define our chosen line-of-sight, we select the location where we wish to calculate the line acceleration, hereafter notated as the ``local" point, with vector position $\ro=\{r_\mathrm{loc},\theta_\mathrm{loc}, \phi_\mathrm{loc}\}$.
The direction for the ray is then given as proceeding in the direction with impact parameter $b$ and locally-centered angle $\phi'$, and the length of the ray from the stellar footpoint to the local point is $\ell$. To define the relation of the locally-centered coordinate system including $\phi'$ to the star-centered spherical coordinate system, we follow \cite{CraOwo95} such that the Cartesian ``primed" coordinate system is given by $\hat{x}'\equiv\hat{r}$, $\hat{y}'\equiv\hat{\theta}$, $\hat{z}'\equiv\hat{\phi}$, and $\phi'$ is given by the standard Cartesian to Spherical coordinate transformation, $\phi'\equiv \mathrm{arctan}(y'/x')$.

\begin{figure*}
\hspace*{\fill}
\begin{subfigure}{0.4\textwidth}
\includegraphics[width=\textwidth]{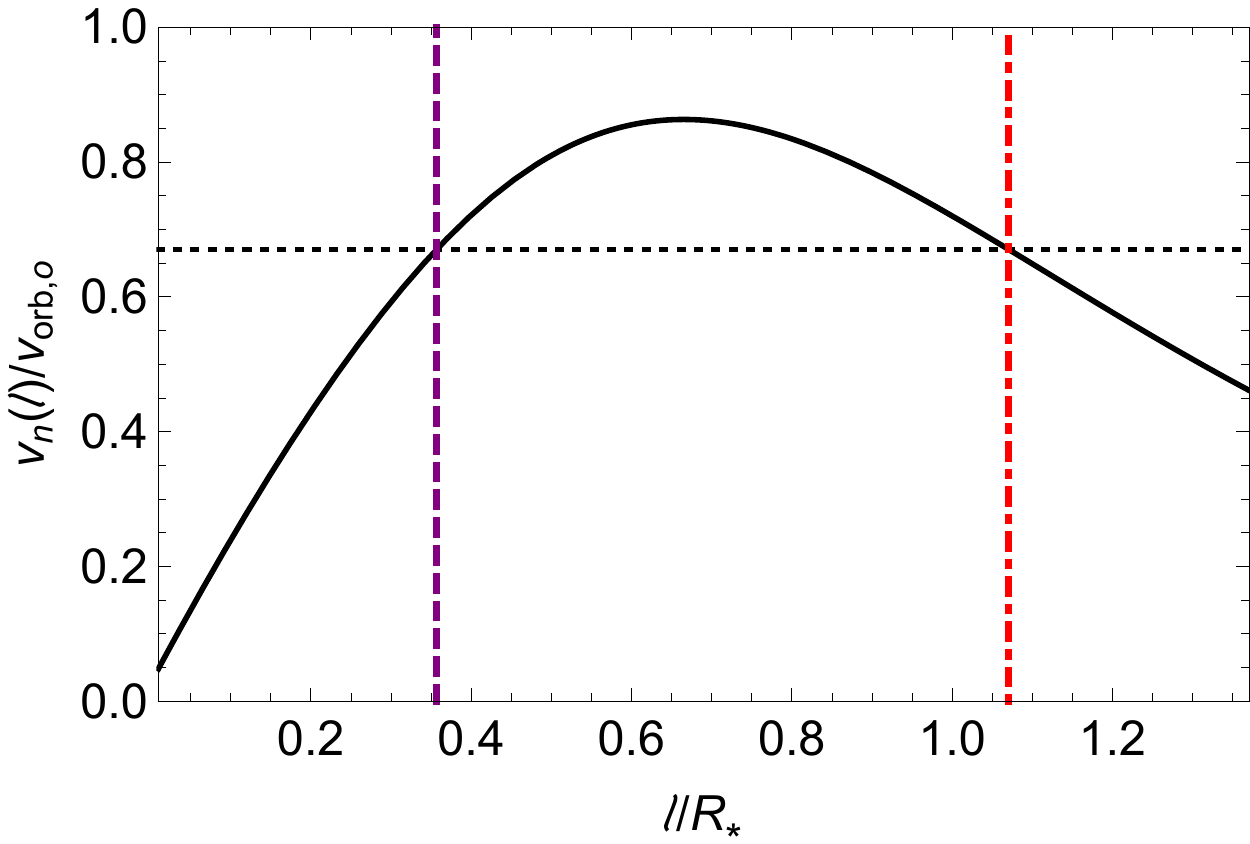}
\caption{$v_n$}
\end{subfigure}\hfill
\begin{subfigure}{0.4\textwidth}
\includegraphics[width=\textwidth]{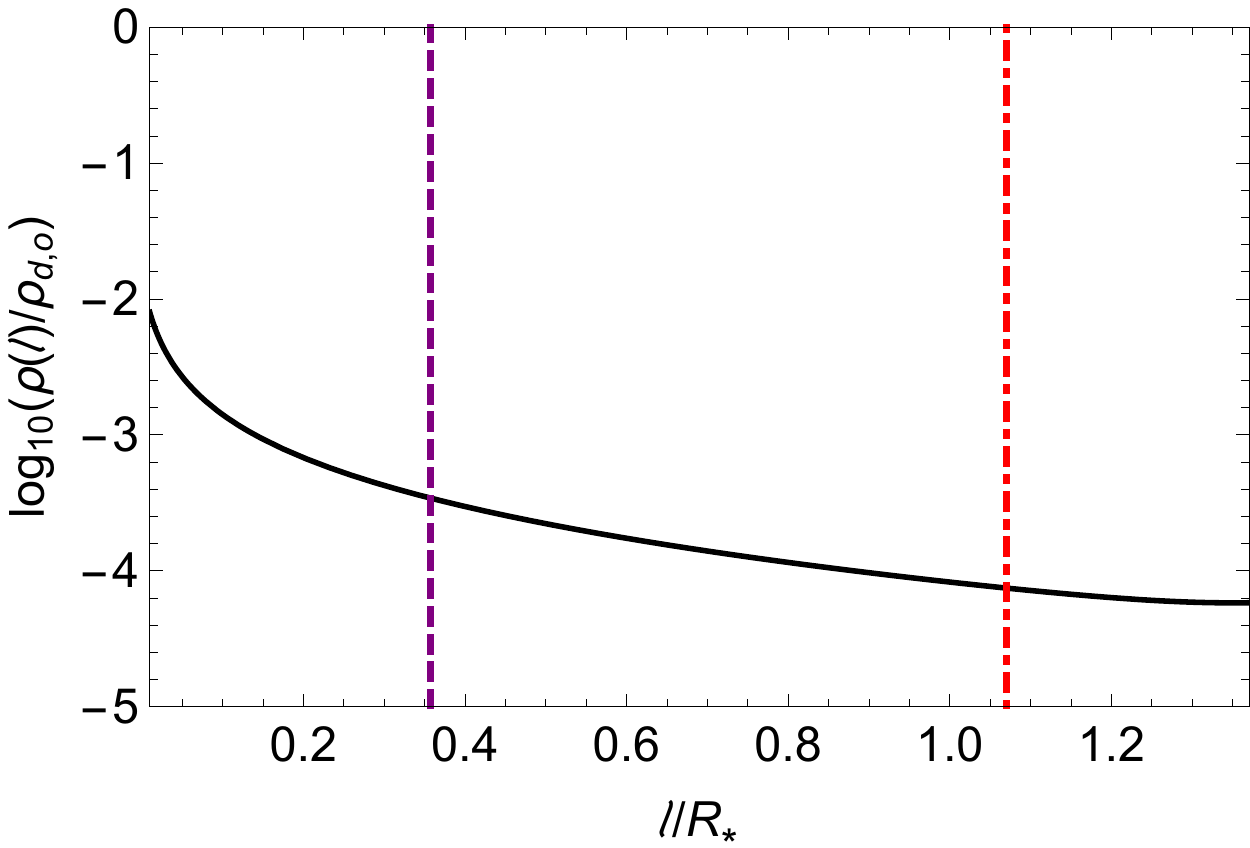}
\caption{$\log_{10}(\rho)$}
\end{subfigure}
\hspace*{\fill}
\caption{The line of sight variation of (a) $v_n$ and (b) $\rho$ for a ray from $r_\mathrm{loc}=2R_\ast$, $\theta_\mathrm{loc}=12^\circ$ to $b=0.5R_\ast$, $\phi'=45^\circ$. The red dashed line indicates the location where we calculate the acceleration and the purple dashed line indicates the location of the inner resonance. Path length along the line of sight is plotted in units of stellar radii with velocity in units of $v_{orb,\mathrm{o}}$ and density in units of $\rho_{d,\mathrm{o}}$.\label{fig:vn_rhon_b0p5}}
\end{figure*}

As an example local point\footnote{We omit $\phi_{\mathrm{loc}}$ as the model is azimuthally symmetric.}, let us select $r_\mathrm{loc}=2R_\ast$, $\theta_\mathrm{loc}=12^\circ$.
If we then look in the direction $b=0.5R_\ast$, $\phi'=45^\circ$, the line-of-sight velocity component is non-monotonic, as can be seen from panel (a) of figure \ref{fig:vn_rhon_b0p5}.
By calling out the local point with a vertical, red, dash-dotted line, we can see that photons Doppler shifted into resonance at the local point also come into resonance approximately $0.35R_\ast$ from the stellar surface, called out by a vertical, purple, dashed line.
Points where photons come into resonance before reaching $\ro$ will be referred to as arising at $\rres$ for the remainder of the paper.
In a case such as this one where $\rres$ is between $\ro$ and the stellar core, photons are scattered out of the line-of-sight, thereby decreasing the intensity seen at the local point.
Note that in the appendix of paper I, we only considered the effects of this type of multiple resonance.

Examining panel (b) of figure \ref{fig:vn_rhon_b0p5} shows that, in addition to velocity, density also varies along the line of sight.
Moreover, while line-of-sight velocity is the same at $\ro$ and $\rres$ by definition, density is not constrained in the same way. 
In paper I, we ignored this variation under the assumption that each resonance was of infinite optical depth.
Since the Sobolev optical depth is proportional to $\tau_s \propto \rho/(dv_n/d\ell)$, where $dv_n/d\ell$ is the velocity gradient along the line-of-sight, it is possible for the optical depths of the resonances at $\ro$ and $\rres$ to be significantly different, a point we return to in section \ref{sec:root-find}.

\begin{figure*}
\hspace*{\fill}
\begin{subfigure}{0.4\textwidth}
\includegraphics[width=\textwidth]{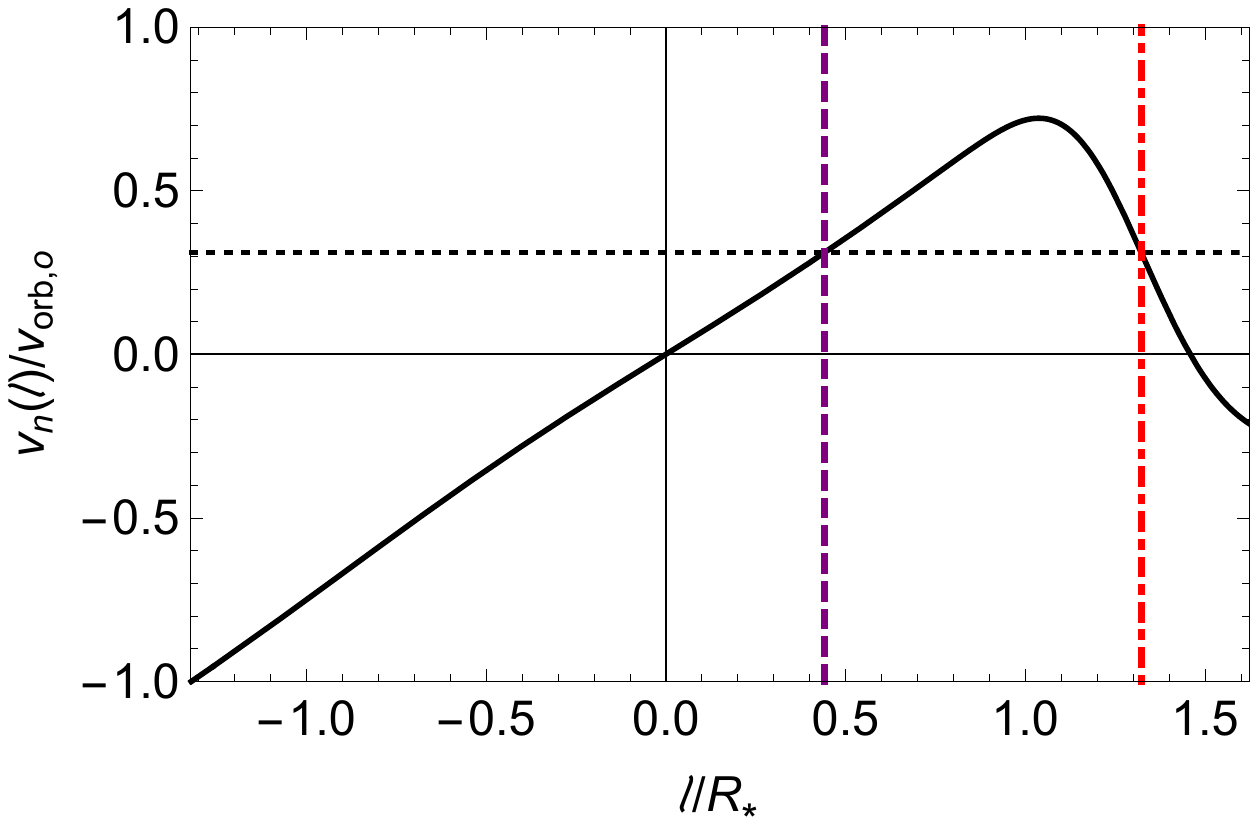}
\caption{$v_n$}
\end{subfigure}\hfill
\begin{subfigure}{0.4\textwidth}
\includegraphics[width=\textwidth]{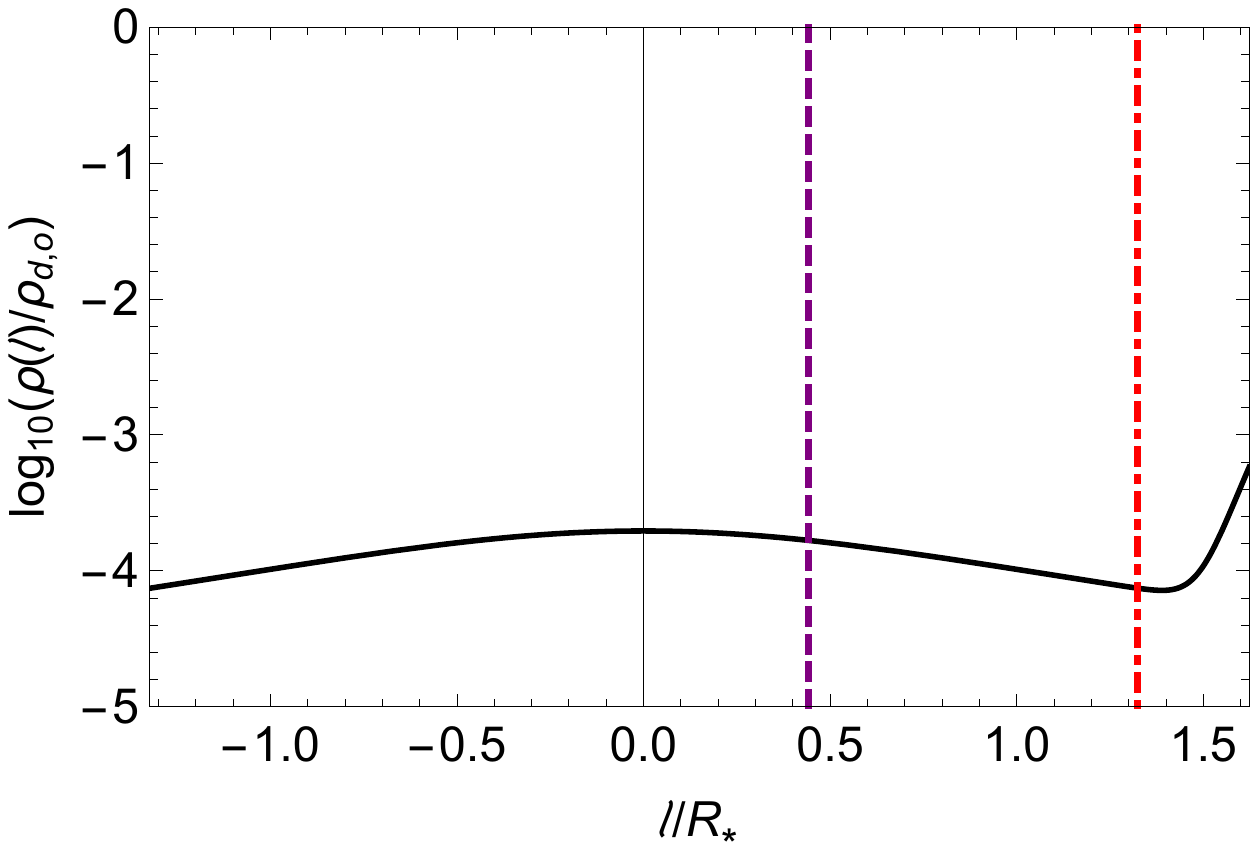}
\caption{$\log_{10}(\rho)$}
\end{subfigure}
\hspace*{\fill}
\caption{As figure \ref{fig:vn_rhon_b0p5}, except now with an off-star ray with $b=1.5R_\ast$.
\label{fig:vn_rhon_b1p5}}
\end{figure*}

Multiple resonances between the local and the stellar core are not the only type that arise.
This is easily demonstrated by again taking $r_\mathrm{loc}=2R_\ast$, $\theta_\mathrm{loc}=12^\circ$ as our local point, but now looking in the direction with $b=1.5 R_\ast$ and $\phi'=45^\circ$.
As confirmed by panel (a) of figure \ref{fig:vn_rhon_b1p5}, we find that here too the non-monotonicity in the line-of-sight velocity causes multiple resonances to arise which, as is shown in panel (b) of figure \ref{fig:vn_rhon_b1p5}, are once again not at the same density.
Since this ray does not impact the stellar core, photons are now scattered \emph{into} the line-of-sight along this ray rather than \emph{out} of the line-of-sight as is the case for multiple resonances between $\ro$ and the stellar core.
This scattering of photons into the line of sight for such ``off-star'' rays can be an important effect for two major reasons, discussed further below:
\begin{enumerate}
\item Multiple resonances between the stellar core and the local point can only decrease the photon flux at the local point\footnote{This ignores the possibility of a population inversion between the local point and the stellar core. If this occurs, stimulated emission can increase the photon flux. For the considered UV driving-lines, however, such a population inversion is highly unlikely.},
 while multiple resonances along rays off the stellar limb can act to ``fill in" some of the lost flux.
\item Off-star multiple resonances allow photons to come from larger impact parameters than the stellar core, most notably at large $r_\mathrm{loc}$ where the stellar core approaches a point source.
\end{enumerate}

While the importance of the filling-in of lost flux is fairly clear, the importance of the arrival direction of photons is potentially less obvious.
To illuminate this issue, recall that the strength of line acceleration is proportional not only to the impinging intensity, but also to the line-of-sight velocity gradient. 
In a purely Keplerian disc, the only velocity gradient along a line-of-sight is the projection of the Keplerian shear onto that line-of-sight.
As $r_\mathrm{loc}$ increases, the decreasing cone angle of the star means that the component of this non-radial velocity gradient along the star-wards lines-of-sight also decreases.
If photons instead arrive from off-star directions, then they also come from angles where the projection of the Keplerian shear is larger, and therefore these ``off-star" photons can be more effective at driving ablation than those coming from the stellar core.
In the following section, we discuss how to quantify these effects, now by calculating the line-acceleration itself as a function of $r$ and $\theta$.

\section{Root-Finding Method}
\label{sec:root-find}

To continue the analysis of paper I, we again use standard root finding methods to locate the position of the inner resonances, $\rres$.
Recall from paper I, and the discussion in the prior section, that we have previously assumed all resonances are of infinite optical depth, such that the intensity along each line of sight can be given by the source function at $\rres$, $I^\mathrm{loc}(\ro)=S(\rres)$.
The introduction of an analytic density structure allows us to relax this approximation, as we can now directly calculate the optical depth at the inner resonance.
For a spectral line with opacity enhancement $q$, the Sobolev \citep{Sob60} optical depth is given by
\begin{equation}
\tau_{s,q}(\mathbf{r})=\frac{\kappa_e q c \rho(\mathbf{r})}{dv_n(\mathbf{r})/d\ell}\;,
\end{equation}
where $dv_n(\mathbf{r})/d\ell$ is the line-of-sight velocity gradient at position $\mathbf{r}$.
Using the formal solution of radiative transfer, the intensity at the ``local" point is then

\begin{equation}\label{eqn:formal_soln}
I^\mathrm{loc}(\ro)=I_{c}e^{-\tau_q(\rres)} + S(\rres)\left(1-e^{-\tau_q(\rres)}\right)\;,
\end{equation}
with $I_c$ the unattenuated core intensity.

By setting this as intensity transmitted through the inner resonance, we can derive the line-acceleration at the local point.
Following the methodology of CAK with the modifications introduced by \cite{OwoCas88} and \cite{Gay95}, we begin by assuming that the distribution of lines in $q$ is given by

\begin{equation}\label{eqn:dNdq}
\frac{dN}{dq}=\frac{\bar{Q}}{\Gamma(\alpha)Q_\mathrm{o}^2}\left(\frac{q}{Q_\mathrm{o}}\right)^{\alpha-2}e^{-q/Q_\mathrm{o}}\;,
\end{equation}
where $\alpha$ is the power-law index introduced by CAK.
In this notation, the CAK power-law distribution is truncated at a maximally strong line of enhancement $Q_\mathrm{o}$ over continuum opacity, while $\bar{Q}$ is the total opacity enhancement in the limit where all the lines are optically thin.
The associated enhancement of line-acceleration over electron scattering is then given by

\begin{equation}\label{eqn:gline_o_ge}
\frac{g_\mathrm{line}(\ro)}{g_e}=q\frac{1-e^{-\tau_q(\ro)}}{\tau_q(\ro)}
\end{equation}
For an ensemble of non-overlapping lines, integration over the line distribution given by equation \ref{eqn:dNdq} and intensity given by equation \ref{eqn:formal_soln} yields, for a single ray in the $\hat{n}$-direction,

\begin{align}\label{eqn:gn}
\mathbf{g}_{n} (\ro,\hat{n}) &=\frac{\kappa_e \bar{Q}[I_c A
+S(\rres)B]}{(1-\alpha)c\tau_\mathrm{o}(\ro)}
\hat{n}\;\\
A &= a - b\\
B &= b + c - a -1\\
a &= (1+\tau_\mathrm{o}(\rres)+\tau_\mathrm{o}(\ro))^{1-\alpha} \\
b &= (1+\tau_\mathrm{o}(\rres))^{1-\alpha}\\
c &= (1+\tau_\mathrm{o}(\ro))^{1-\alpha}\;,
\end{align}
where $\tau_\mathrm{o}$ is $\tau_q$ for $q=Q_\mathrm{o}$.
As is also the case in the absence of multiple resonances, we then must integrate $g_{n}$ over solid angle to recover the 3-dimensional line-acceleration, namely $\mathbf{g}_\mathrm{lines}=\oint \mathbf{g}_{n}d\Omega$.

In principle, the source function used in equations \ref{eqn:formal_soln} and \ref{eqn:gn} requires a self-consistent treatment of all the non-local couplings implied by the existence of multiple resonances.
For the purposes of this analysis, we take the simplifying approximation of an analytic source function.
As one limiting case, we consider the optically thin scattering source function, given by the geometric dilution of mean intensity from the stellar core as

\begin{equation}\label{eqn:Sthin}
S_\mathrm{thin}(r_\mathrm{res})\equiv I_c\frac{1-\mu_\ast(r_\mathrm{res})}{2}\;,
\end{equation}
where $\mu_\ast(r_\mathrm{res})\equiv cos(\theta_\ast)=\sqrt{1-(R_\ast/r_\mathrm{res})^2}$ is the cosine of the angular size of the star as seen from $r_\mathrm{res}$.

For the other limiting case of an optically thick scattering source function, we use \citep{OwoRyb85},

\begin{equation}\label{eqn:Sthick}
S_\mathrm{thick}(r_\mathrm{res})\equiv S_\mathrm{thin}(r_\mathrm{res})\left(\frac{1+\sigma(r_\mathrm{res}) E_3(r_\mathrm{res})}{1+\sigma(r_\mathrm{res})/3}\right)\;,
\end{equation}
where we use the notation

\begin{equation}
E_k(r)\equiv \frac{1}{k}\frac{1-\mu_\ast(r)^k}{1-\mu_\ast(r)}=\frac{1}{k} \sum_{i=1}^k \mu_\ast(r)^{k-1}\;.
\end{equation}
Equation \ref{eqn:Sthick} also uses the definition \citep{Cas74},

\begin{equation}\label{eqn:Cas_sig}
\sigma(r)= \frac{d \ln v(r)}{d \ln r}-1\;.
\end{equation}
As the vast majority of the inner resonances occur in the wind, we take the simplifying approximation that $v(r)$ in equation \ref{eqn:Cas_sig} is given only by the radial wind velocity such that $\sigma(r)=(2R_\ast-r)/(r-R_\ast)$, thereby making both $S_\mathrm{thin}$ and $S_\mathrm{thick}$ analytic functions of only radius.

Even with the simplification of an analytic source function, the procedure of calculating line-acceleration from multiple resonances described here is too costly to undertake at every time step for a numerical hydrodynamics simulation.
For a single analytic velocity and density structure, however, it is quite tractable.
The next section presents the results of just such an analysis, carried out on the density and velocity structures presented in section \ref{sec:mult_res}.

\section{Results}
\label{sec:results}

\begin{figure*}
\begin{subfigure}{\textwidth}
\includegraphics[width=\textwidth]{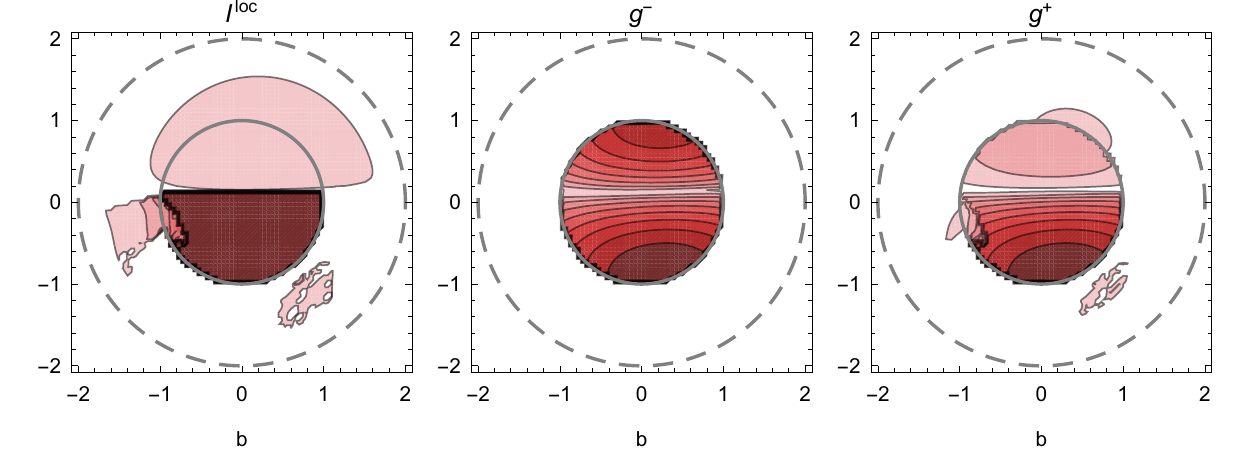}
\end{subfigure}

\begin{subfigure}{0.45\textwidth}
\includegraphics[width=\textwidth]{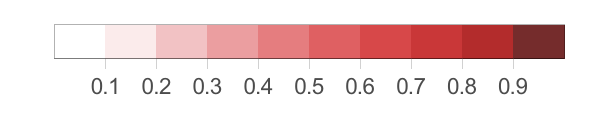}
\end{subfigure}
\caption{One point maps of (from left to right) intensity $I^{loc}$, radial line acceleration without resonances  $g^-(\hat{n})$, and radial line acceleration accounting for resonances $g^+(\hat{n})$. This figure uses the source function $S_\mathrm{thin}$. Intensity has been normalized by $I_c$, while both force maps are normalized by Max[$g^-(\hat{n})$] such that all panels use the same colorbar.\label{fig:I_g_gs_map}}
\end{figure*}

In parallel with the analysis in the appendix of paper I, we again begin by considering the conditions at a single point with fixed $\ro$.
With our additional consideration of off-star rays, our maps now continue out to lines-of-sight perpendicular to the direction toward the center of the star (i.e. $b=r_\mathrm{loc}$).
Additionally, by using equation \ref{eqn:gn}, we now include maps of the magnitude of the contribution from each ray to the radial line-acceleration in the case where we ignore multiple resonances (hereafter $g^-$) and in the case where we account for multiple resonances (hereafter $g^+$).
For $S_\mathrm{thin}$ and $r_\mathrm{loc}=2R_\ast$, $\theta_\mathrm{loc}=12^\circ$, figure \ref{fig:I_g_gs_map} shows maps of $g^-$ and $g^+$, as well as a map of intensity accounting for multiple resonances, $I^{loc}$.
The solid grey circles denote the angular extent of the stellar core, and dashed grey circles denote $b=r_\mathrm{loc}$.

\begin{figure}
\includegraphics[width=0.49\textwidth]{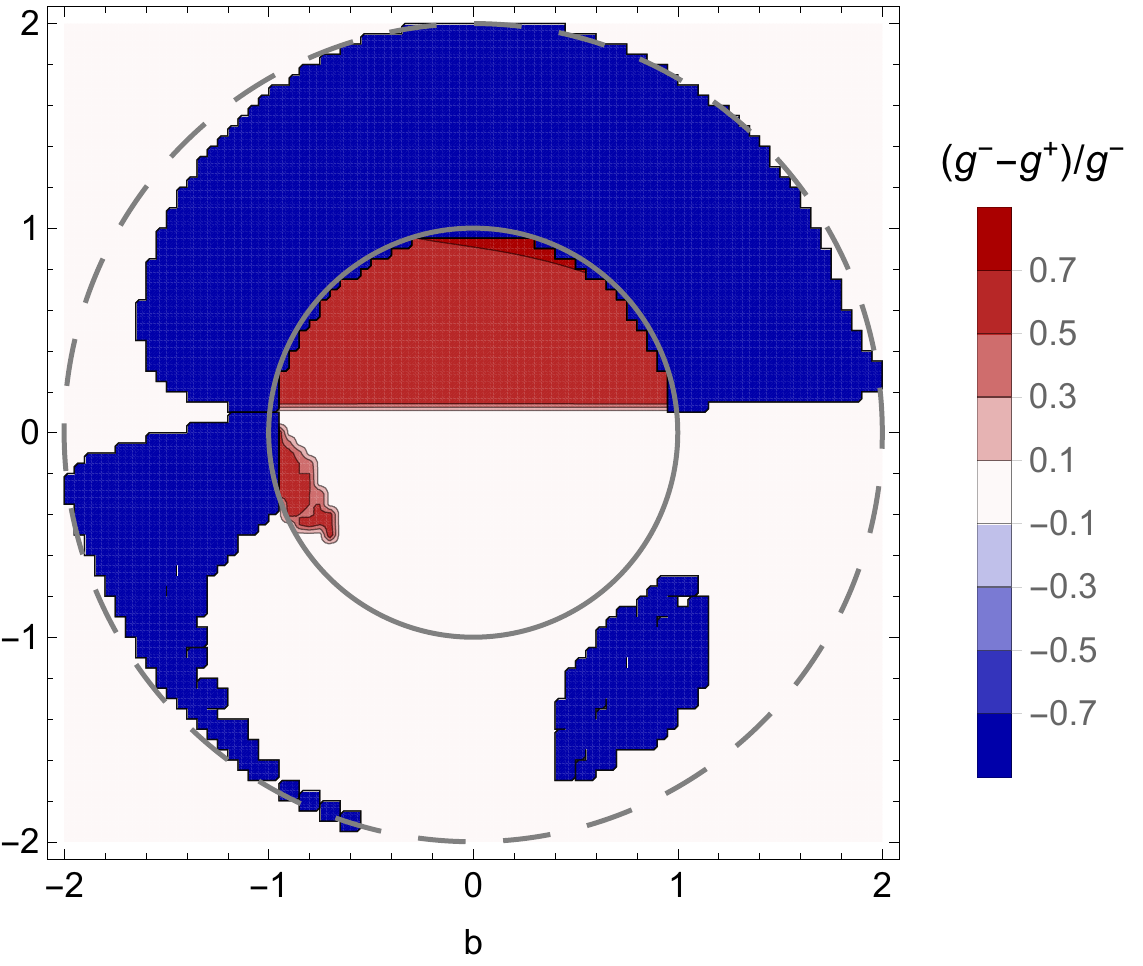}
\caption{One point map of $(g^--g^+)/g^-$. The gray solid and dashed circles represent the angular extent of the star and the maximum impact parameter for which $g^+$ is calculated respectively.\label{fig:grat_map}}
\end{figure}

As discussed in section \ref{sec:mult_res} and above, both $I^{loc}$ and $g^+$ are no longer constrained to the stellar core.
This can be illustrated by plotting the relative ratio of accelerations without and with multiple resonances $(g^- - g^+)/g^-$, as is done in figure \ref{fig:grat_map}.
Here, it is helpful to mentally divide the figure into two regions: $b \leq R_\ast$ and $b>R_\ast$.
For $b \leq R_\ast$, $g^+$ is always less than or equal to $g^-$, consistent with all multiple resonances on these rays scattering photons out of the line-of-sight.
On the other hand, for $b>R_\ast$ the opposite is true, consistent now with all multiple resonances on these rays scattering photons \emph{into} the line-of-sight.
Because $g^-$ is defined to be zero for $b>R_\ast$, any non-zero $g^+$ in this region will mean $(g^- - g^+)/g^-=-\infty$ and thus the blue region in figure \ref{fig:grat_map} can only indicate which rays have multiple resonances, not the strength of the line driving along these rays.
Also because of the angular confinement of $g^-$ to on-star rays, the white regions take on slightly different meanings on and off star.
For $b \leq R_\ast$, white indicates the region where the local point receives unaltered stellar line-driving, while for $b>R_\ast$ white indicates the region where there is still no line-driving despite accounting for multiple resonances.

\begin{figure*}
\hspace*{\fill}
\begin{subfigure}{0.45\textwidth}
\includegraphics[width=\textwidth]{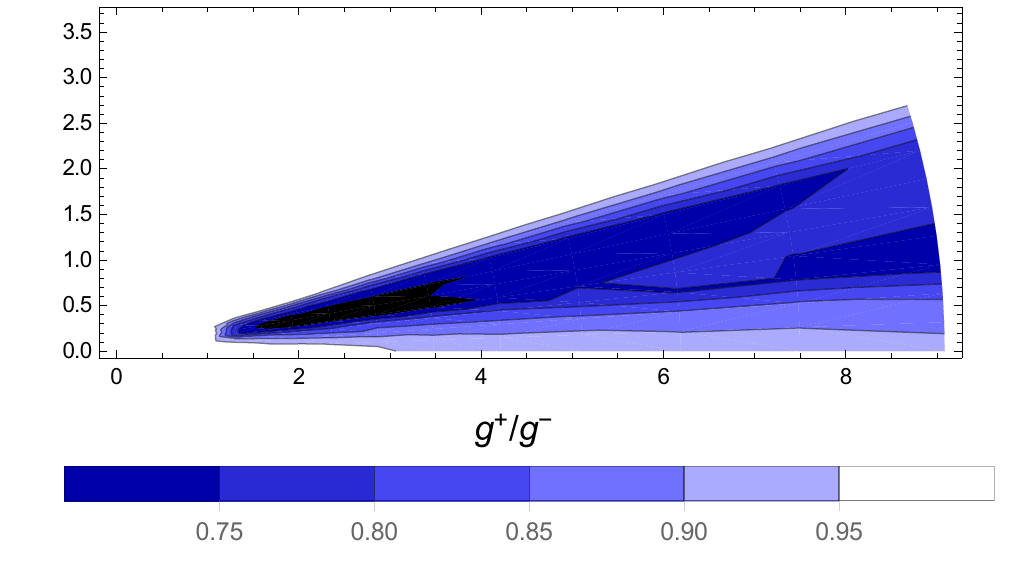}
\caption{$S_\mathrm{thin}, 0\leq b \leq R_\ast$}
\end{subfigure}\hfill
\begin{subfigure}{0.45\textwidth}
\includegraphics[width=\textwidth]{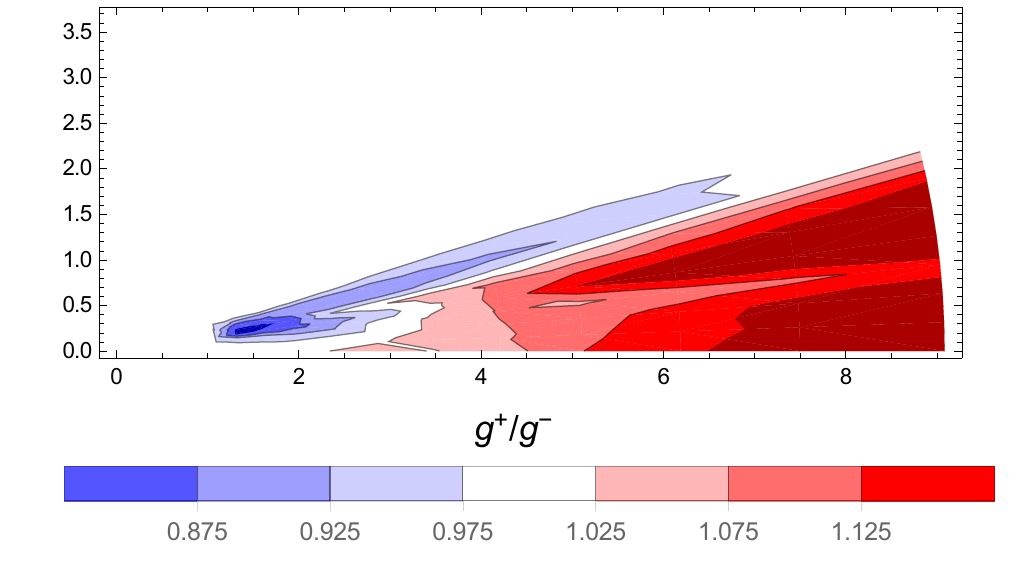}
\caption{$S_\mathrm{thin}, 0\leq b \leq r_\mathrm{o}$}
\end{subfigure}
\hspace*{\fill}

\hspace*{\fill}
\begin{subfigure}{0.45\textwidth}
\includegraphics[width=\textwidth]{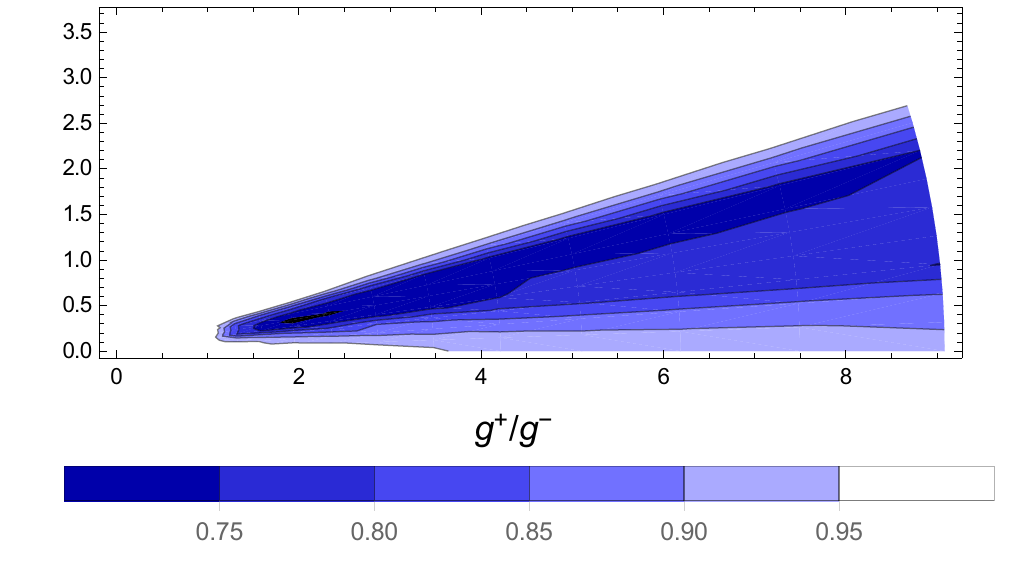}
\caption{$S_\mathrm{thick}, 0\leq b \leq R_\ast$}
\end{subfigure}\hfill
\begin{subfigure}{0.45\textwidth}
\includegraphics[width=\textwidth]{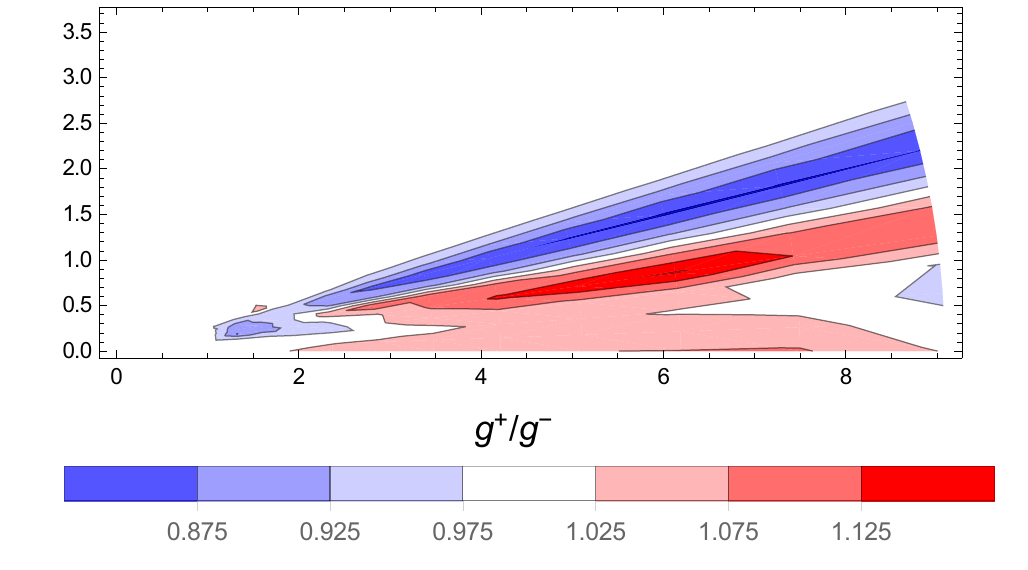}
\caption{$S_\mathrm{thick}, 0\leq b \leq r_\mathrm{o}$}
\end{subfigure}
\hspace*{\fill}

\caption{Contour map of $g^+/g^-$ for (top) $S_\mathrm{thin}$ and (bottom) $S_\mathrm{thick}$  both (left) omitting and (right) including off star rays.\label{fig:g+_over_g-}}
\end{figure*}

If we now integrate $g^-$ and $g^+$ over the star-wards $2\pi$ steradians for a range of $\ro$, and take their ratio, we arrive at maps that show the overall modification of line-driving from the inclusion of multiple resonances.
In panels (a) and (b) of figure \ref{fig:g+_over_g-}, we show the results of this analysis for $S_\mathrm{thin}$.
In order to more directly assess the importance of off-star rays, we have included a panel that only accounts for line-driving from the stellar core, as well as a panel for the integration over all $b \leq r_\mathrm{o}$.
As becomes immediately evident, the inclusion of off-star rays raises the value of $g^+$ everywhere inside the disc.
In fact, in the subsurface layers of the disc, the effect can even be strong enough to cause a net \emph{enhancement} of $g^+$ over $g^-$.
However, as $g^-$ is only a small fraction of the stellar gravity in these deeper layers, this increase is of academic interest more than it is impactful to the net line-driving. 
In the ablation layers, where changes in line-acceleration can have more significant dynamical effects, we find that nowhere does $g^+$ drop below $\sim 80\%\; g^-$.

For comparison, panels (c) and (d) of figure \ref{fig:g+_over_g-} plot the same quantities, now calculated with $S_\mathrm{thick}$.
For the case where we omit off star rays (c), the change from using $S_\mathrm{thin}$ is relatively modest, in that the spatial distribution of reduction in line-acceleration is quite similar and the minimum of $g^+/g^-$ is only a few percent larger in the case with $S_\mathrm{thick}$ than with $S_\mathrm{thin}$.
The change is somewhat more pronounced when we account for off star rays (d).
Notable in the morphology is the extension of the region of force reduction along the entire surface layer of the disc.
The range of reductions for both cases ($\lesssim 20\%$), however, suggests that multiple resonances do not strongly impact ablation in either the case with $S_\mathrm{thin}$ or $S_\mathrm{thick}$.
This result is even more promising when we examine the region within the first stellar radius above the surface, where paper I infers that the majority of ablation is launched.
When using $S_\mathrm{thin}$, the maximum reduction in acceleration ($\sim 20\%$) occurs here.
However, in the case with $S_\mathrm{thick}$, this reduction is only $\sim 13\%$.
Since we expect the real source function to be somewhere between $S_\mathrm{thin}$ and $S_\mathrm{thick}$, we also should expect the reduction of acceleration in this critical region to never exceed a modest $20\%$.

Independent of the source function used, there are two important points from this analysis that should be emphasized.
First, for understanding the effect of multiple resonances on radiative ablation, calculating the line-acceleration instead of the flux reduction is more accurate.
As discussed already, from figure \ref{fig:g+_over_g-} we can see that calculating the modification of line-acceleration also gives us a more optimistic picture for how radiative ablation will behave in the presence of multiple resonances than calculations of flux reductions.
Considering the case with only on-star rays and $S_\mathrm{thin}$, the appendix of paper I shows $\sim 50\%$ flux reduction over a large portion of the ablation layer.
Meanwhile, the same case with only on-star rays and $S_\mathrm{thin}$ here shows reductions of only $\sim 30-35\%$ in line-acceleration over this same region, emphasizing that multiple resonances cause line-acceleration to no longer be proportional to local flux.
Second, for both source functions used, the inclusion of off-star rays does indeed act to ``fill-in" some of the lost acceleration along on-star rays, yielding the aforementioned $\lesssim20\%$ reduction in acceleration in all cases.

\section{Summary and Future Work}
\label{sec:conclusions}

The study presented here uses the newly derived equation \ref{eqn:gn} to quantify the effects on line-acceleration of multiple resonances arising in analytic density and velocity structures.
A key result is the difference between the $\sim50\%$ reductions inferred for flux in the appendix of paper I and the $15-20\%$ reductions in line-acceleration inferred here.
Due to the scale of this effect, we determine that the role of multiple resonances in modifying line-driven ablation is small enough to likely be ignorable in future analyses.
This is especially true in light of the potential scale of other effects that have as of yet gone unexplored.

One such effect is the modification of line-driven ablation by the continuum optical depth of the circumstellar disc.
Thus far, we have only considered discs for which $\tau_d \lesssim 1$.
However, discs around Classical Be stars can reach continuum optical depths above unity in the equatorial plane \citep[see, e.g.][]{RivCar13}.
As well, discs around still forming stars can easily reach mean continuum optical depths many orders of magnitude higher in the equatorial plane (see, e.g. \cite{KuiYor13}, their fig. 5; \cite{KuiYor15}, their fig. 16).
Future models of both these cases therefore require a model for continuum optical depth effects; particularly, one that can be employed in a numerical hydrodynamics code.

Another potentially important issue in treating line-driven ablation of Classical Be star discs is the so called ``mass reservoir effect".
First introduced in the context of the viscous decretion disc model by \cite{GhoCar17}, the central idea of the mass reservoir effect is that the finite duration of the disc building process does not allow for the organization of disc material into the pure $\rho\propto r^{-3.5}$ power-law profile used as the initial conditions for the simulations in paper I.
For the simulations of \cite{GhoCar17}, this different mass profile means that the effective $\alpha\sim 0.2$ required to destroy Classical Be star discs is substantially lower than previously inferred $\alpha\sim 1$.
For simulations like those in paper I, it is not yet clear what effect redistributing the mass will have on the ablation process.

Finally, more work should still be dedicated to exploring the role of multiple resonances.
For the problem of radiative ablation, there are two questions in particular that should still be addressed. 
First, what will the effect of changing the line-driving strength be on the asymptotic hydrodynamic structure we infer from ablation?
Particularly, is the analytic structure we used in this analysis consistent with the force adjustments implied by the analysis?
To investigate this, future tests should iterate the process of determining an analytic fit to the asymptotic hydrodynamic structure, determining force corrections from this analytic fit, and then applying these to the line-acceleration in the hydrodynamic code to get a new asymptotic structure to analyze for force corrections.
A second question of particular interest is the accuracy of the source functions used in this analysis.
As discussed in section \ref{sec:root-find}, calculating the source function requires a self-consistent accounting for the non-local couplings implied by the presence of multiple resonances.
Here we took the limiting cases of $S_\mathrm{thin}$ and $S_\mathrm{thick}$ to bracket the problem, but a self-consistent treatment of multiple resonances requires a self-consistent source function.
Future work will analyze this problem using such a source function.

\section*{Acknowledgements}

This study was conducted within the Emmy Noether research group on ``Accretion Flows and Feedback in Realistic Models of Massive Star Formation" funded by the German Research Foundation under grant no. KU 2849/3-1. Additionally, the authors would like to thank the anonymous referee for comments which helped clarify the discussion found here.

\bibliographystyle{mn2e}
\bibliography{biblio}

\end{document}